\documentclass[%
aps,
pre,
superscriptaddress,
tightenlines,
showpacs,showkeys,
a4paper,
12pt,
reprint,
notitlepage
]{revtex4-1}
\usepackage[english]{babel}
\usepackage{amssymb,amsmath,stmaryrd,array}

\usepackage{graphicx}
\usepackage{subfig}

\makeatletter

\newcommand{\sca}[2]{\ensuremath{\bigl({#1}\cdot{#2}\bigr)}}
\newcommand{\avr}[1]{\ensuremath{\langle{#1}\rangle}}

\newcommand{\mum}{$\mu$m}
\newcommand{\dega}{\ensuremath{^\circ}}
\newcommand{\degc}{$^\circ$C}
\newcommand{\myarrow}[1]{\ensuremath{\xrightarrow{{#1}^\circ\mathrm{C}}}}

 \newcommand{\bs}[1]{\boldsymbol{#1}}
 \newcommand{\vc}[1]{\mathbf{#1}}
 \newcommand{\mvc}[1]{\mathbf{#1}}
 \newcommand{\uvc}[1]{\hat{\mathbf{#1}}}
 
 \newcommand{\ind}[1]{\mathrm{#1}}

\newcommand{\dd}{\mathrm{d}}

\newcommand{\eff}{\mathrm{eff}}

\newcommand{\vac}{\mathrm{vac}}
\newcommand{\med}{\mathrm{m}}

\makeatother

\begin{document}
\DeclareGraphicsExtensions{.eps,.png,.pdf}
\title{
Light modulation in planar aligned
short-pitch deformed-helix ferroelectric liquid crystals  
}

 \author{Svetlana~P.~Kotova}
 \email[Email address: ]{kotova@fian.smr.ru}
\affiliation{%
Lebedev Physical Institute,
Leninsky Prospekt 53, 119991 Moscow, Russia
 }
\affiliation{%
Samara State Aerospace University (SSAU), Moskovskoe shosse 34, 443086 Samara, Russia
 }

\author{Sergey~A.~Samagin}
 \email[Email address: ]{samagin@fian.smr.ru}
\affiliation{%
Lebedev Physical Institute,
Leninsky Prospekt 53, 119991 Moscow, Russia
 }


\author{Evgeny~P.~Pozhidaev}
 \email[Email address: ]{epozhidaev@mail.ru}
\affiliation{%
Lebedev Physical Institute,
Leninsky Prospekt 53, 119991 Moscow, Russia
 }
\affiliation{%
 Hong Kong University of Science and Technology,
 Clear Water Bay, Kowloon, Hong Kong
 }

\author{Alexei~D.~Kiselev}
\email[Email address: ]{alexei.d.kiselev@gmail.com}
\affiliation{%
 Hong Kong University of Science and Technology,
 Clear Water Bay, Kowloon, Hong Kong
 }
\affiliation{%
 Saint Petersburg National Research University of Information Technologies,
 Mechanics and Optics (ITMO University),
 Kronverskyi Prospekt 49,
 197101 Saint Petersburg, Russia}

\date{\today}

\begin{abstract}
We study both experimentally and theoretically
modulation of light in
a planar aligned
deformed-helix ferroelectric liquid crystal (DHFLC) cell
with subwavelength helix pitch,
which is also known as a short-pitch DHFLC.
In our experiments,
azimuthal angle of the in-plane optical axis
and electrically controlled parts of
the principal in-plane refractive indices
were measured as a function of voltage applied across the cell.
Theoretical results giving the effective optical tensor
of a short-pitch DHFLC expressed in terms of
the smectic tilt angle and the refractive indices 
of FLC
are used to fit the experimental data.
Optical anisotropy of the FLC material is found
to be weakly biaxial.
For 
both the transmissive and reflective modes,
the results of fitting are applied
to model phase and amplitude modulation of
light in the DHFLC cell.
We demonstrate
that,
if the thickness of the DHFLC layer is about $50$~\mum,
the detrimental effect of
field-induced rotation
of the in-plane optical axes
on 
the characteristics of 
an axicon designed
using the DHFLC spatial light modulator
in the reflective mode
is negligible. 
\end{abstract}

\pacs{%
61.30.Gd, 78.20.Jq, 42.70.Df, 
42.79.Kr, 42.79.Hp 
}
\keywords{%
helix deformed ferroelectric liquid crystal; 
subwavelength pitch;
modulation of light; axicon.
}
 \maketitle

\section{Introduction}
\label{sec:intro}

High-speed, low power consuming light modulation 
is in high demand
for a variety of photonic devices
used as building blocks of displays and
optical information processors. 
These include
tunable lenses, focusers,
wavefront correctors and 
correlators~\cite{Wilkinson:applopt:1995,Kotova:optex:2002,Kotova:jopt:2003,Xu:optex:2012,Hu:optex:2004,Moreno:optexp:2012}.

For the most part of such devices, 
in addition to fast switching times, 
it is of crucial importance
to have a $2\pi$ modulation so that 
the phase can be smoothly tuned from
zero to $2\pi$.
Liquid crystal (LC) spatial light modulators (SLMs) are 
widely used as devices to modulate amplitude,
phase, or polarization of light waves
in space and time~\cite{Efron:bk:1995}. 
In LC-SLMs, nematic liquid crystals
are among the most popular LC phases.
However, nematics are known to have slow
response time and, in addition, this slow response gets worse 
if the LC layer thickness increases in
order to obtain the $2\pi$ phase modulation.  
Therefore, many efforts are in progress to optimize
the various LC electro-optical modes for the high speed light
modulation.

\begin{figure*}[!tbh]
\centering
\subfloat[Smectic cone]{
  \resizebox{65mm}{!}{\includegraphics*{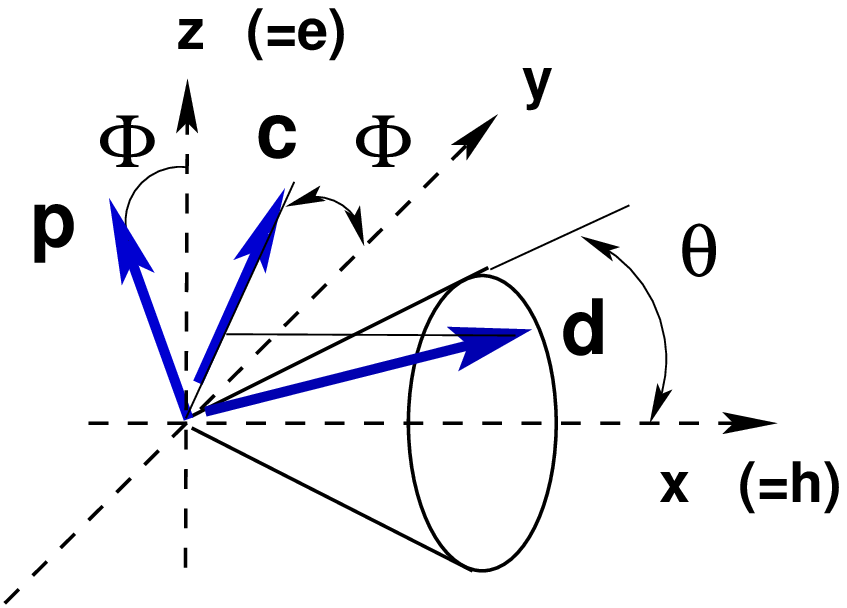}}
\label{subfig:director}
}
\subfloat[Planar aligned FLC cell]{
  \resizebox{70mm}{!}{\includegraphics*{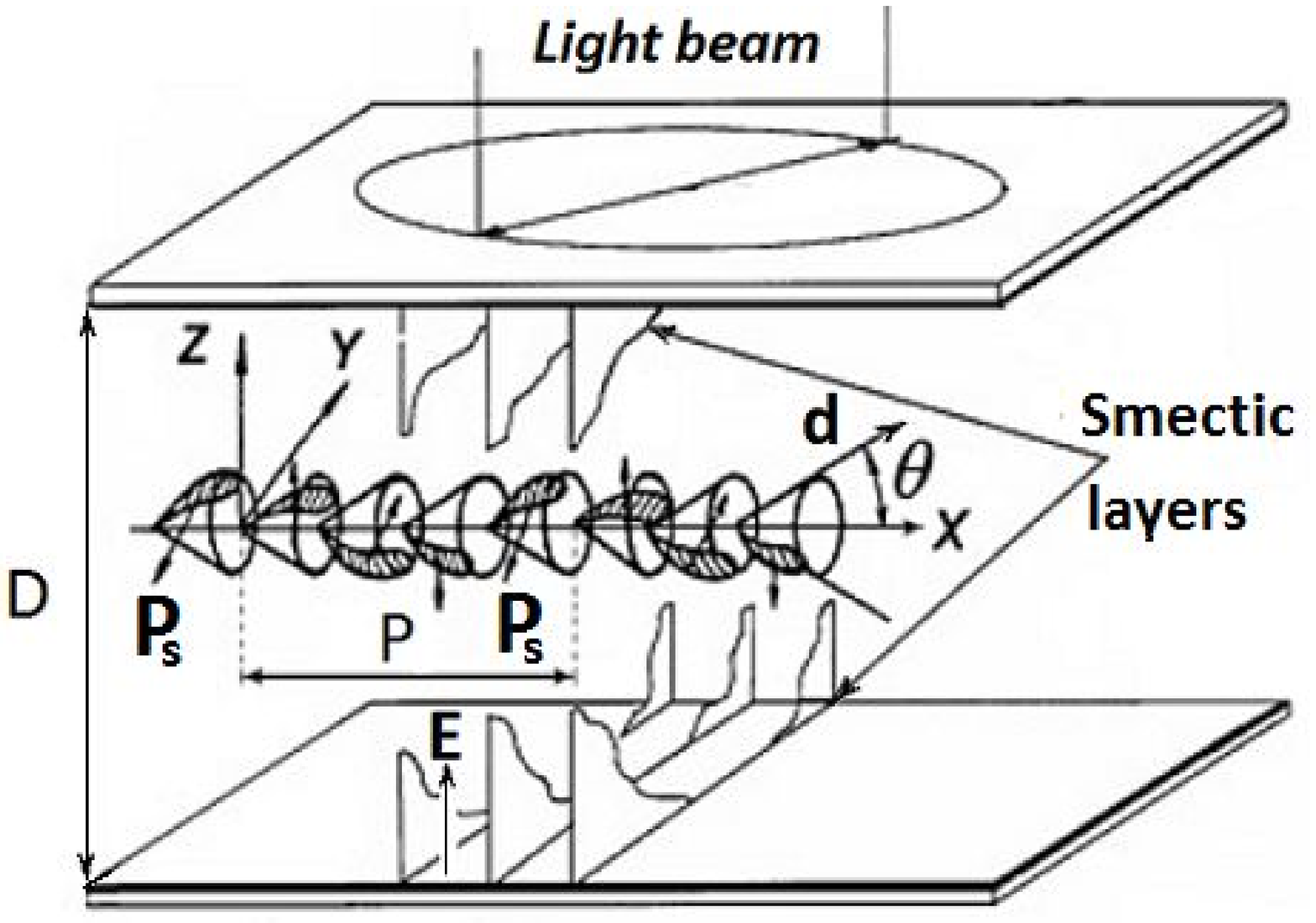}}
\label{subfig:cell}
}
\caption{%
(Color online)
  Geometry of (a)~smectic cone and (b)~planar aligned FLC cell with
uniform lying helix. 
}
\label{fig:geom}
\end{figure*}

Ferroelectric liquid crystals
(FLCs) represent an alternative
and most promising chiral liquid crystal material 
which is characterized by very fast response time
(a detailed description of FLCs can be found, e.g.,
in monographs~\cite{Lagerwall:bk:1999,Oswald:bk:2006}).
Equilibrium orientational structures in FLCs
can be described as helical twisting patterns
where FLC molecules align on average along
a local unit director
\begin{align}
&
\uvc{d}=
\cos\theta\,\uvc{h}+
\sin\theta\,\uvc{c},
\label{eq:director}
  \end{align}
where $\theta$ is the smectic tilt angle; 
$\uvc{h}$ is the twisting axis normal to the smectic layers and
$\uvc{c}\perp\uvc{h}$ is the $c$-director.
The FLC director~\eqref{eq:director}
lies on the smectic cone 
depicted in Fig.~\ref{subfig:director}
with
the \textit{smectic tilt angle} $\theta$
and rotates
in a helical fashion about a uniform twisting axis
$\uvc{h}$ forming the FLC helix with 
the \textit{helix pitch}, $P$.
This rotation is described by
the azimuthal angle
around the cone $\Phi$
that specifies
orientation of the $c$-director in the plane perpendicular to
$\uvc{h}$ and depends on
the dimensionless coordinate along the twisting axis
\begin{align}
&
\phi=2 \pi \sca{\uvc{h}}{\vc{r}}/P= q x,
\label{eq:phi}    
 \end{align}
where $q=2\pi/P$ is the helix twist wave number. 

Figure~\ref{fig:geom}
illustrates
the important case of a \textit{uniform lying} FLC helix
in the slab geometry 
with the smectic layers normal to the substrates
and
\begin{align}
&
\uvc{h}=\uvc{x},
\quad
\uvc{c}=\cos\Phi\,\uvc{y}+
\sin\Phi\,\uvc{z},
\quad
\vc{E}=E\,\uvc{z},
\label{eq:director-planar}    
  \end{align}
where $\vc{E}$ is the applied electric field
which is linearly coupled to
the \textit{spontaneous ferroelectric polarization}
\begin{align}
&
\vc{P}_s=P_s\uvc{p},
\quad
\uvc{p}=
\uvc{h}\times\uvc{c}=\cos\Phi\,\uvc{z}-
\sin\Phi\,\uvc{y},
\label{eq:pol-vector}
  \end{align}
where $\uvc{p}$ is
the \textit{polarization unit vector}.

This  is the geometry of surface stabilized FLCs
(SSFLCs)
pioneered by Clark and Lagerwall in Ref.~\cite{Clark:apl:1980}
where they studied electro-optic response of 
FLC cells confined between two parallel plates
subject to homogeneous boundary conditions
and made thin enough to suppress the bulk FLC 
helix. 
Figure~\ref{subfig:cell} also describes
the geometry of 
deformed helix FLCs (DHFLCs)
as it was originally introduced in~\cite{Beresnev:lc:1989}.
The approach to
light modulation that uses 
the electro-optical properties
of helical structures in 
DHFLCs with subwavelength pitch also known as 
the \textit{short-pitch DHFLCs}
will be of our primary concern.

In short-pitch DHFLC cells, the FLC helix
is characterized by
a submicron helix pitch, $P<1$~\mum, 
and a relatively large tilt angle, 
$\theta>30$\dega. 
Electro-optical response of DHFLC cells
exhibits a  number of peculiarities that make them
useful for LC devices such as
high speed spatial light 
modulators~\cite{Abdul:mclc:1991,Cohen:aplopt:1997,Pozhidaev:ferro:2000,Kiselev:pre:2013,Kiselev:ol:2014}, 
colour-sequential liquid crystal display cells~\cite{Pozhidaev:jsid:2012}
and optic fiber sensors~\cite{Kiselev:photsens:2012,Brodzeli:jlighttech:2013}.
The effects caused by electric-field-induced distortions of the helical structure 
underline the mode of operation of such cells. 
In a typical experimental setup, these effects
are probed by performing measurements of 
the transmittance of normally incident light
through a cell placed between crossed polarizers.

In this article, our goal is to examine
light modulation in
\textit{planar aligned} short-pitch DHFLC (PADHFLC) cells
with uniform lying FLC helix 
(the twisting axis is parallel to the substrates)
and the related physical characteristics.
This is the case that was studied theoretically
in Refs~\cite{Kiselev:pre:2011,Kiselev:pre:2:2014}
where 
the transfer matrix approach to polarization
gratings was employed to define
the effective dielectric tensor of short-pitch 
DHFLCs.
In particular, it was found that,
by contrast
to the case of vertically aligned DHFLCs~\cite{Kiselev:pre:2013,Kiselev:ol:2014}, 
the in-plane optical axes of PADHFLCs sweep in the
plane of the cell under the action of  applied electric field
thus producing changes in the polarization state of the incident
light.
More generally,
biaxial anisotropy and rotation of the optical axes
induced by the electric field
in short-pitch DHFLC cells
can be interpreted as 
the \textit{orientational Kerr effect}~\cite{Kiselev:pre:2013,Kiselev:ol:2014,Kiselev:pre:2:2014}.

For detailed experimental characterization
of this effect, we employ 
the experimental technique
based on the Mach-Zehnder interferometer
that goes beyond
the limitations of 
the above mentioned standard experimental 
procedure and provides additional information
on the principal refractive indices.
Then we use the results as input parameters 
to study
light modulation in the DHFLC cell
operating in both the transmissive and reflective modes.
Our investigation into the effects of amplitude
modulation is based on the results of modelling
of an \textit{axicon} designed
using the DHFLC spatial light modulator
in the reflective mode.

An axicon as a cylindrically symmetric optical element 
that transforms an incident plane wave 
into a narrow beam of light along the optical axis
have a long history
dating back to the original papers by 
John~N.~McLeod~\cite{McLeod:josa:1954,McLeod:josa:1960}.
Laser beams propagating
through axicons have two significant properties: 
(a)~they
generate a line focus, where the on-axis intensity stays high over
much longer distances compared to focusing by conventional
lenses; and (b)~they generate ring-like intensity profiles in the far
field. 
Both of these properties proved useful in many applications
such as atom guiding and trapping~\cite{Manek:optcomm:1998},
annular focusing in laser machining~\cite{Belanger:aplopt:1978},
generation of quasi-non-diffracting Bessel-like
beams~\cite{Monk:optcomm:1999,Arlt:optcomm:2000,Ismail:jopt:2012},
subdiffraction limit imaging~\cite{Snoeyink:ol:2013}
and optical
micromanipulation~\cite{Arlt:optcomm:2001,Kotova:qe:2014,Kampmann:aplopt:2014}.

The layout of the paper is as follows.
In Sec.~\ref{sec:kerr-effect}
we introduce our notations,
describe the recent theoretical results~\cite{Kiselev:pre:2:2014}
on the effective optical tensor of short-pitch biaxial FLCs, 
and discuss the orientational Kerr effect.
Experimental details are given in Sec.~\ref{sec:expt},
where we describe the samples and the setup employed to perform
measurements.
The experimental data are fitted using
the expression for the effective optical tensor.
Sec.~\ref{sec:modulation} deals with modulation of light
in the DHFLC cells in both transmissive and reflective modes. 
Experimental results are used to compute
the amplitudes and phases of the components of
transmitted and reflected light waves.
DHFLC spatial light modulator
acting as an axicon
is modelled as
a two-dimensional (2D) array of pixels
which are DHFLC cells.
In order to study the effects of amplitude modulation,
the intensity distribution
in the focal plane of the axicon is evaluated
for both DHFLC and ideal (no amplitude modulation)
axicons.
Finally, in Sec.~\ref{sec:discussion} we draw the results together and
make some concluding remarks.
Technical results are relegated to Appendices.

\section{Orientational Kerr  effect}
\label{sec:kerr-effect}

In this section, we introduce notations and briefly discuss
the electro-optical properties of short-pitch DHFLC cells
described by the effective dielectric (optical) tensor,
$\bs{\varepsilon}_{\eff}$,
defined in terms of averages over distorted FLC helical 
structure~\cite{Kiselev:pre:2011} (see Appendix~\ref{sec:diel-tensor-dhf}).
For the geometry of uniform lying FLC helix (see Fig.~\ref{subfig:cell}),
we recapitulate the analytical results for the optical tensor
of a biaxial ferroelectric liquid crystal with
subwavelength pitch~\cite{Kiselev:pre:2:2014}.
In the subsequent
section,
these results will be used to interpret
the experimental data. 

We consider a FLC layer of thickness $D$
with the $z$ axis 
which, as is indicated in Fig.~\ref{fig:geom}, 
is normal to the bounding surfaces: $z=0$ and $z=D$,
and introduce
the effective  dielectric tensor,
$\bs{\varepsilon}_{\eff}$,
describing a homogenized DHFLC helical structure.

For a biaxial FLC, the components of the dielectric tensor,
$\bs{\varepsilon}$,
are given by
\begin{align}
&
\epsilon_{i j}=
  \epsilon_{\perp}
\delta_{i j}+
(\epsilon_{1}-\epsilon_{\perp})\,
d_i d_j
+
(\epsilon_{2}-\epsilon_{\perp})\,
p_i p_j
\notag
\\
&
=
  \epsilon_{\perp}
(
\delta_{i j}+u_1 d_i d_j
+u_2 p_i p_j
),
\label{eq:diel-tensor}
\end{align}
where 
$i,j\in\{x,y,z\}$,
$\delta_{ij}$ is the Kronecker symbol;
$d_i$ ($p_i$) is the $i$th component
of the FLC director (unit polarization vector)
given by Eq.~\eqref{eq:director} (Eq.~\eqref{eq:pol-vector});
$u_i=(\epsilon_{i}-\epsilon_{\perp})/\epsilon_{\perp}=\Delta\epsilon_i/\epsilon_{\perp}=r_i-1$
are the \textit{anisotropy parameters}
and $r_1=\epsilon_1/\epsilon_\perp$ ($r_2=\epsilon_2/\epsilon_\perp$)
is the \textit{anisotropy (biaxiality) ratio}.
Note that, in the case of uniaxial anisotropy with $u_2=0$,
the principal values of the dielectric tensor are:
$\epsilon_2=\epsilon_{\perp}$ and
$\epsilon_{1}=\epsilon_{\parallel}$,
where $n_{\perp}=\sqrt{\mu\epsilon_{\perp}}$
($n_{\parallel}=\sqrt{\mu\epsilon_{\parallel}}$)
is the ordinary (extraordinary) refractive index
and the magnetic tensor of FLC 
is assumed to be isotropic with the magnetic permittivity $\mu$. 
We shall also assume that 
the medium surrounding the layer is optically
isotropic and is characterized by 
the dielectric constant $\epsilon_{\med}$,
the magnetic permittivity $\mu_{\med}$ and the refractive index
$n_{\med}=\sqrt{\mu_{\med} \epsilon_{\med}}$.

Assuming that
the pitch-to-wavelength ratio $P/\lambda$
is sufficiently small,
the effective dielectric tensor can be expressed
in terms of the averages over the pitch of
distorted FLC helical structure.
Explicit formulas for the components of the tensor
are given in Appendix~\ref{sec:diel-tensor-dhf}.
These formulas 
can be used to derive the effective optical tensor of
homogenized short-pitch DHFLC cell for both vertically and planar aligned
FLC helix~\cite{Kiselev:pre:2013,Kiselev:pre:2:2014}.

We concentrate on the geometry
of planar aligned DHFLC helix
shown in Fig.~\ref{subfig:cell}.
For this geometry, 
the effective dielectric tensor can be written
in the following form~\cite{Kiselev:pre:2:2014}:
\begin{align}
  \label{eq:eff-diel-planar}
  &
\bs{\varepsilon}_{\eff}=
\begin{pmatrix}
  \epsilon_h+\gamma_{xx}\alpha_E^2 & \gamma_{xy} \alpha_E& 0\\
  \gamma_{xy} \alpha_E,& \epsilon_p+\gamma_{yy}\alpha_E^2 & 0\\
0 & 0 & \epsilon_p-\gamma_{yy}\alpha_E^2
\end{pmatrix}.
\end{align}
The \textit{zero-field dielectric constants}, 
$\epsilon_h$ and $\epsilon_p$,
that enter the tensor~\eqref{eq:eff-diel-planar} are given by
\begin{subequations}
  \label{eq:epsilon_ph}
\begin{align}
&
\label{eq:epsilon_h}
\epsilon_h
/\epsilon_{\perp}
=(n_h/n_{\perp})^2=
r_2^{-1/2}
\biggl\{
\sqrt{r_2}
\notag
\\
&
+u_1 \cos^2\theta
\left(
\frac{r_2-1}{\sqrt{u}+\sqrt{r_2}}+u^{-1/2}
\right)
\biggr\},
\\
&
\label{eq:epsilon_p}
\epsilon_p/\epsilon_{\perp}=
(n_p/n_{\perp})^2=
\sqrt{r_2 u},
\\
&
\label{eq:u}
u= r_2 (v+1)=
u_1\sin^2\theta+1.
\end{align}
\end{subequations}
A similar result for 
the \textit{coupling coefficients}
$\gamma_{xx}$,
$\gamma_{yy}$ and $\gamma_{xy}$
reads
\begin{subequations}
  \label{eq:coupling-coeffs}
   \begin{align}
&
\label{eq:gxx-u}
\gamma_{xx}/\epsilon_{\perp}=
\frac{3\sqrt{r_2/u}}{(\sqrt{u}+\sqrt{r_2})^2}
(u_1\cos\theta\sin\theta)^2,
\\
&
\label{eq:gyy-u}
\gamma_{yy}/\epsilon_{\perp}=
\frac{3\sqrt{r_2u}}{(\sqrt{u}+\sqrt{r_2})^2}
(u-r_2),
\\
&
\label{eq:gxy-u}
\gamma_{xy}/\epsilon_{\perp}=
\frac{2\sqrt{r_2}}{\sqrt{u}+\sqrt{r_2}}
u_1\cos\theta\sin\theta.
    \end{align}
\end{subequations}

Note that, following Ref.~\cite{Kiselev:pre:2013},
we have 
introduced the electric field parameter
\begin{align}
  \label{eq:alpha_E}
\alpha_E=\chi_E E/P_s,  
\end{align}
where $\chi_E=\partial \avr{P_z}/\partial E$ 
is
the dielectric susceptibility of 
the Goldstone mode~\cite{Carlsson:pra:1990,Urbanc:ferro:1991}
and $P_z=P_s\cos\Phi$.

Note that the simplest averaging procedure 
previously used in Refs.~\cite{Abdul:mclc:1991,Kiselev:pre:2011,Kiselev:pre:2013} 
heavily relies on the first-order approximation where the director
distortions are described by the term linearly proportional to the
electric field.
Quantitatively, the difficulty with this approach is that
the linear approximation may not be suffice
for accurate computing of the second-order contributions
to the diagonal elements of the dielectric
tensor~\eqref{eq:eff-diel-planar}.
In Ref.~\cite{Kiselev:pre:2:2014}, 
the results~\eqref{eq:eff-diel-planar}--\eqref{eq:gxy-u}
were derived by using
the modified averaging technique that 
allows high-order corrections to the dielectric tensor 
to be accurately estimated
and
improves agreement between the theory and the experimental
data in the high-field region.

The dielectric tensor~\eqref{eq:eff-diel-planar}
is characterized by the three generally different principal values
(eigenvalues)
and the corresponding optical axes (eigenvectors)
as follows
\begin{align}
&
  \label{eq:eff-diel-diag-planar}
  \bs{\varepsilon}_{\eff}=
\epsilon_z \uvc{z}\otimes\uvc{z}
+\epsilon_{+} \uvc{d}_{+}\otimes\uvc{d}_{+}
+\epsilon_{-} \uvc{d}_{-}\otimes\uvc{d}_{-},
\\
&
\label{eq:epsilon_z}
\epsilon_{z}=n_{z}^{\,2}=\epsilon_{zz}^{(\eff)}=\epsilon_p-\gamma_{yy}\alpha_E^2,
\\
&
\label{eq:epsilon_pm}
\epsilon_{\pm}=n_{\pm}^{\,2}=\bar{\epsilon}\pm\sqrt{[\Delta\epsilon]^2+[\gamma_{xy} \alpha_{E}]^2}
\end{align}
where
\begin{align}
&
  \label{eq:epsilon_avr}
  \bar{\epsilon}=(\epsilon_{xx}^{(\eff)}+\epsilon_{yy}^{(\eff)})/2
=\bar{\epsilon}_0+(\gamma_{xx}+\gamma_{yy})\alpha_{E}^2/2,
\\
&
  \label{eq:Delta_epsilon}
 \Delta\epsilon=(\epsilon_{xx}^{(\eff)}-\epsilon_{yy}^{(\eff)})/2=
\Delta\epsilon_0+(\gamma_{xx}-\gamma_{yy})\alpha_{E}^2/2,
\\
&
  \label{eq:Delta_epsilon0}
\bar{\epsilon}_0=(\epsilon_{h}+\epsilon_{p})/2,\:
\Delta\epsilon_0=(\epsilon_{h}-\epsilon_{p})/2.
\end{align}
The in-plane optical axes are given by
\begin{align}
&
  \label{eq:d_plus}
  \uvc{d}_{+}=\cos\psi_d\,\uvc{x}+
\sin\psi_d\,\uvc{y},
\quad
  \uvc{d}_{-}=\uvc{z}\times  \uvc{d}_{+},
\\
&
  \label{eq:psi_d}
2\psi_\dd =\arg[\Delta\epsilon +i \gamma_{xy} \alpha_E].
\end{align}

From Eq.~\eqref{eq:eff-diel-planar}, it is clear that
the zero-field dielectric tensor is
uniaxially anisotropic with the optical axis directed along the
twisting axis $\uvc{h}=\uvc{x}$. 
The applied electric field
changes the principal values 
(see Eqs.~\eqref{eq:epsilon_z} and~\eqref{eq:epsilon_pm})
so that 
the electric-field-induced anisotropy is generally biaxial.
In addition, the in-plane principal optical axes are rotated about the
vector of electric field, $\vc{E}\parallel \uvc{z}$,
by the angle $\psi_\dd$ given in Eq.~\eqref{eq:d_plus}.

In the low electric field  region,
the electrically induced part of the principal values
is typically dominated by the
Kerr-like nonlinear terms proportional to $E^2$
\begin{subequations}
\begin{align}
&
  \label{eq:n_plus_approx}
  n_{+}\approx n_h+\frac{1}{2n_h}
\left\{
\gamma_{xx}+\frac{\gamma_{xy}^2}{n_h^2-n_p^2}
\right\}\alpha_E^2,
\\
&
  \label{eq:n_minus_approx}
  n_{-}\approx n_p+\frac{1}{2n_p}
\left\{
\gamma_{yy}-\frac{\gamma_{xy}^2}{n_h^2-n_p^2}
\right\}\alpha_E^2,
\end{align}
\end{subequations}
whereas 
the electric field dependence of the angle $\psi_\dd$ is
approximately linear: $\psi_\dd\propto E$.
This effect is caused by
the electrically induced distortions of the helical
structure and bears some resemblance to the electro-optic Kerr effect. 
Following Refs.~\cite{Kiselev:pre:2013,Kiselev:ol:2014}, 
it will be referred to as the \textit{orientational Kerr effect}.

It should be emphasized that this effect
differs from
the well-known Kerr effect which is  a quadratic electro-optic effect
related to the electrically induced birefringence
in optically isotropic (and transparent) materials~\cite{Weinberger:pml:2008}.
 Similar to 
polymer stabilized blue phase liquid crystals~\cite{Yan:apl:2010,Yan:apl:2013},
it is governed by the effective dielectric tensor
of a nanostructured chiral smectic liquid crystal.
This tensor is defined
through averaging over the FLC orientational structure
(see Appendix~\ref{sec:diel-tensor-dhf}).

\section{Experiment}
\label{sec:expt}

In this section we present the experimental results
on the principal refractive indices and orientation of the
in-plane optical axis
measured as a function of the applied electric field
in DHFLC cells.

\subsection{Material and sample preparation}
\label{subsec:flc-sample}

In our experiments we used 
the FLC mixture  FLC-587 
(from
P. N. Lebedev Physical Institute of Russian Academy of Sciences)
 as a material for the DHFLC layer.
The FLC-587 is an eutectic mixture of the three compounds
whose chemical structures 
are described in Ref.~\cite{Kiselev:pre:2013}.  
The phase transitions sequence of this FLC during heating
up from the solid crystalline phase
is:
Cr$\myarrow{+12}$SmC$^{\star}\myarrow{+110}$SmA$^{\star}\myarrow{+127}$Is,
while cooling from smectic C$^\star$ phase crystallization occurs
around -10$^\circ$C~ -15$^\circ$C. 
The spontaneous polarization, $P_s$, 
and the helix pitch, $P$, at room temperature (22$^\circ$C)
are 150~nC/cm$^2$ and 
$150$~nm, respectively. 

The cell was sandwiched between two glass
substrates covered by ITO (indium tin oxide) 
and aligning films of the 
thickness $20$~nm, and
the gap was fixed by spacers at $D=50$~mm.
Geometry of the cells is schematically depicted in Fig~\ref{subfig:cell}. 

\begin{figure}[!htb]
\centering
  \resizebox{60mm}{!}{\includegraphics*{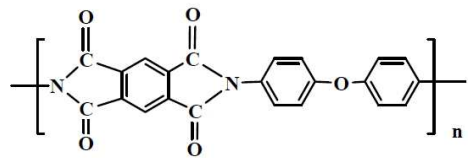}}
\caption{%
Chemical structure of the compound forming the aligning layer.
}
\label{fig:chem}
\end{figure}

High quality planar alignment 
yielding the contrast ratio
up to $200:1$ 
was achieved 
using
4,4'-oxydianiline dianhydride (PMDA-ODA) as aligning layers. Chemical
formula of this polyimide after imidization is shown in Fig.~\ref{fig:chem}. 

PMDA-ODA dissolved in dimethyl-formamide
(the concentration was about 0.2 wt\%)
was spin-coated onto the ITO surface. 
The polyimide film then was dried on 
the ITO substrate for 30-40~min at
temperature 180\degc, 
and subsequent imidization was done at temperature
within the interval 
275\degc--290\degc\  for about 1~h. 

Following the method of Ref.~\cite{Zhukov:cryst:2006}, 
after cooling down, the polyimide films were
rubbed with a cotton shred to provide aligning layers anisotropy.
The FLC mixture was then injected into the cells in the isotropic phase 
by capillary action.
 
Our task was to obtain
regular helix alignment in the cell with the helix axis parallel to the
glass plates.  
For this purpose,
the FLC cell was subjected to 
an additional electrical training 
with a square-wave function of maximum
field amplitude ranged from 5~V/\mum\  to 9~V/\mum\ and 
the frequency in the range 
between 0.5~Hz and 2~kHz~\cite{Pozhidaev:lc:2:2010}. 
The obtained alignment 
was inspected by observing 
textures within the cell in a
polarizing microscope.

\begin{figure*}[!tbh]
\centering
  \resizebox{95mm}{!}{\includegraphics*{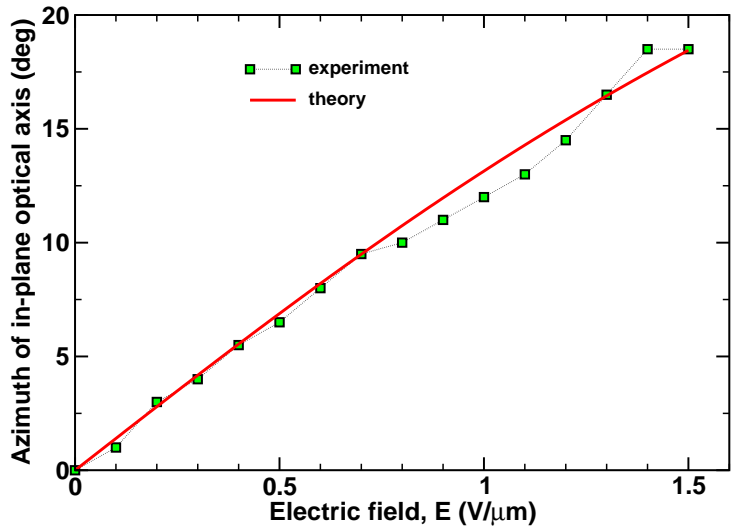}}
\caption{%
(Color online)
Azimuthal angle $\psi_{\dd}$ of the major in-plane optical axis
as a function of electric field 
for the DHFLC cell of thickness $D=50$~\mum\
filled with the FLC mixture FLC-587~\cite{Kiselev:pre:2013}.
The experimental points are marked by squares.
Solid line represents
the theoretical curve
computed from Eq.~\eqref{eq:d_plus} 
with $P_s/\chi_E\approx 4.83$~V/\mum.
Parameters of the mixture are: 
$n_{\perp}\approx 1.52$ ($\epsilon_{\perp}\approx 2.3$)
is the ordinary refractive index;
 $n_{\parallel}\approx 1.77$
($\epsilon_{\parallel}\equiv\epsilon_1\approx 3.13$) 
is the extraordinary refractive index, 
$\theta=35.5$~deg is the tilt angle
and $r_2=1.05$ is the biaxiality ratio.
}
\label{fig:phi_expt_th}
\end{figure*}

\subsection{Measurement of azimuthal angle of in-plane optical axis}
\label{subsec:opt-axis-azimuth}

In our experiments, 
we have used a low power He-Ne laser
($\lambda=632.8$~nm)
as a light source.
Initially, without applied voltage,
the FLC cell was placed between the crossed polarizers
and rotated so as to minimize the transmission
of normally incident light.
Then the cell was subjected to the time varying voltage
of the symmetric square-wave form with the frequency
of $40$~Hz and the amplitude ranged from
zero to $100$~V.
 Under the action of the applied electric field,
the in-plane optical axis rotates
about the normal to the substrates
(the $z$ axis)
and its azimuthal angle $\psi_{\dd}$ changes.

The angle $\psi_{\dd}$ characterizing
the electric-field-induced in-plane reorientation 
of the optical axis 
was measured by rotating
the cell around the $z$ axis
and detecting the angle where 
the intensity of the transmitted light
at positive voltages is minimal. 
The experimental results
for this angle obtained at different
values of the voltage amplitude 
are presented in Fig.~\ref{fig:phi_expt_th}.

\subsection{Measurement of principal in-plane refractive indices}
\label{subsec:refract-indx}

In order to perform measurements of
the principal values of the in-plane refractive indices,
$n_{+}$ and $n_{-}$ (see Eq.~\eqref{eq:epsilon_pm}), 
we have used
the well-known experimental method 
which is based on
a Mach-Zehnder two-arm interferometer
(it is detailed in many textbooks such as~\cite{Born:bk:1999}).

\begin{figure*}[!tbh]
\centering
  \resizebox{160mm}{!}{\includegraphics*{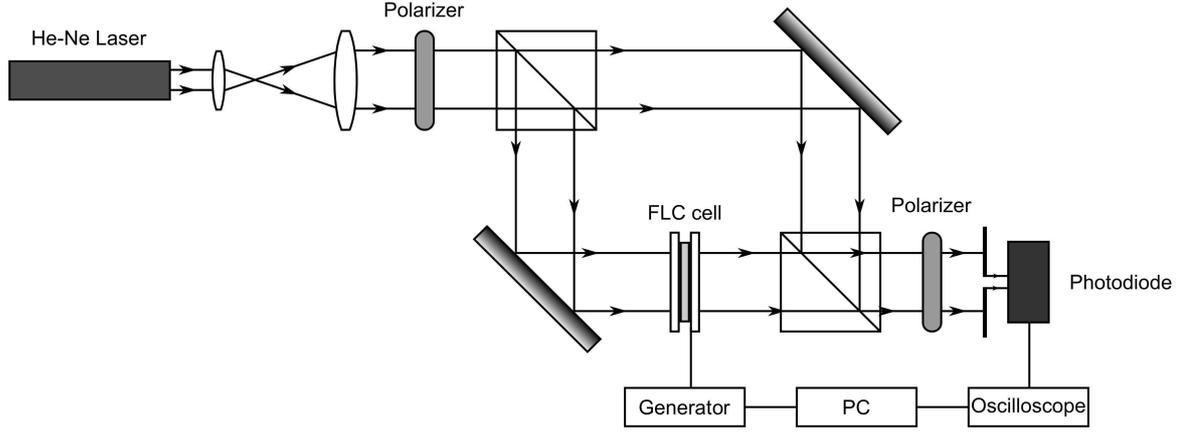}}
\caption{%
Experimental setup for measurements of electrically
controlled parts of the in-plane refractive indices
of the DHFLC cell.
}
\label{fig:setup}
\end{figure*}

In this method,
as it can be seen in Fig.~\ref{fig:setup},
a beam splitter divides a linear polarized incident light
passed through the input polarizer
into two paths and
the FLC cell is placed in the path of
the sample beam.
The sample and reference beams are then recombined
and pass through the output polarizer so that
the interfering beams after the polarizer
are collected by a photodiode.

\begin{figure*}[!tbh]
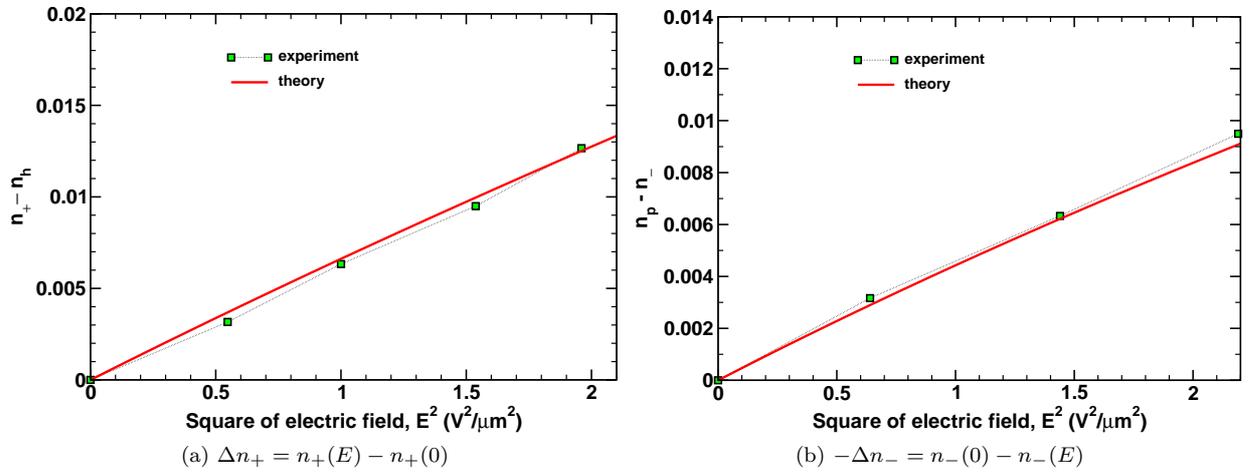

\centering
\subfloat[$\Delta n_{+}=n_{+}(E)-n_{+}(0)$]{
  \resizebox{80mm}{!}{\includegraphics*{fig5a.eps}}
\label{subfig:dn_plus}
}
\subfloat[$-\Delta n_{-}=n_{-}(0)-n_{-}(E)$]{
  \resizebox{80mm}{!}{\includegraphics*{fig5b.eps}}
\label{subfig:dn_minus}
}
\caption{%
(Color online)
Electrically controlled differences 
of the principal in-plane refractive indices
(a)~$\Delta n_{+}$ and (b)~$-\Delta n_{-}$ 
as a function of applied electric field. 
}
\label{fig:delta-n}
\end{figure*}

Given the amplitude of the voltage and
the corresponding value of
the principal axis azimuthal angle $\psi_{\dd}$,
the polarizers were rotated so that 
the polarization vector of the incident light
is either parallel or perpendicular to
the optical axis.
In both cases, measurements 
giving the electrically controlled part of
the corresponding refractive index
were performed
during the half-period of positive applied voltage.
The results are shown in Fig.~\ref{fig:delta-n}. 

\subsection{Results}
\label{subsec:results}

There are three optical characteristics of
the DHFLC cell that 
we have measured in our experiments:
the azimuthal angle of the optical axis
$\psi_\dd$ and the electrically controlled
parts of two principal refractive indices,
$\Delta n_{\pm}=n_{\pm}(E)-n_{\pm}(0)$.
The experimental data for 
electric field dependence of
the principal axis angle
and $\Delta n_{\pm}$ are presented
in Fig~\ref{fig:phi_expt_th} and Fig.~\ref{fig:delta-n},
respectively.

We can now use formulas
for $n_{\pm}$ [see Eq.~\eqref{eq:epsilon_pm}] 
and $\psi_\dd$ [see Eq.~\eqref{eq:psi_d}]
to fit the experimental data.
For this purpose, we assume that
the FLC mixture is characterized by the parameters:
$\epsilon_{\perp}\approx 2.3$ ($n_{\perp}\equiv n_o\approx 1.52$), 
$\epsilon_{\parallel}=3.13$ ($n_{\parallel}\equiv n_e\approx 1.77$)
and $\theta=35.5$\dega.
The theoretical curves 
shown in Figs.~\ref{fig:phi_expt_th}-\ref{fig:delta-n}
are computed at
$P_s/\chi_E\approx 4.83$~V/\mum\ 
and $r_2=\epsilon_2/\epsilon_{\perp}=1.05$.
Interestingly, the value of the biaxiality ratio
differs from unity and thus optical anisotropy of
the mixture appears to be weakly biaxial.

\begin{figure*}[!tbh]
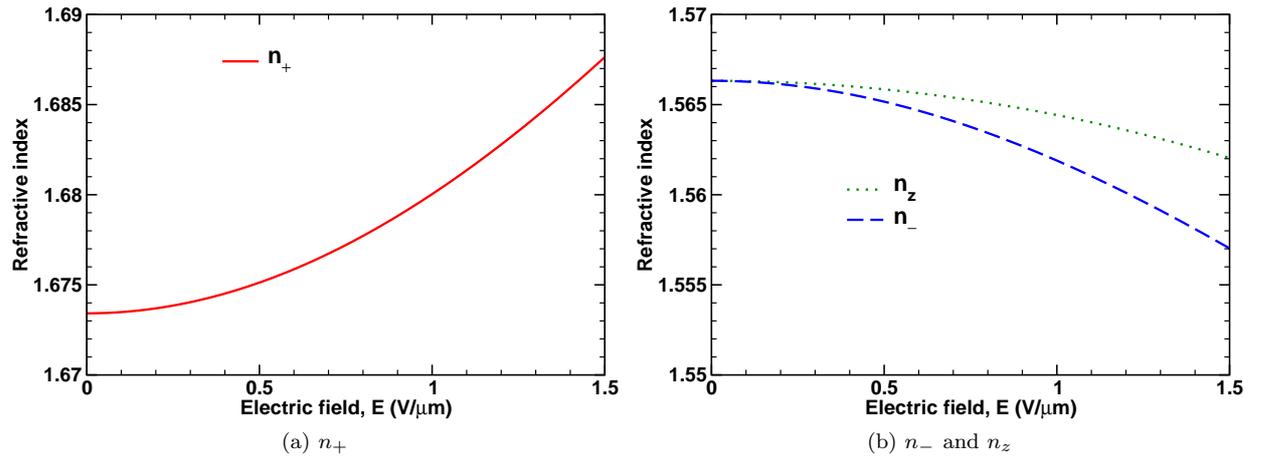

\centering
\subfloat[$n_{+}$]{
  \resizebox{80mm}{!}{\includegraphics*{fig6a.eps}}
\label{subfig:n_pl}
}
\subfloat[$n_{-}$ and $n_{z}$]{
  \resizebox{80mm}{!}{\includegraphics*{fig6b.eps}}
\label{subfig:n_mz}
}
\caption{%
(Color online)
Electric field dependence of principal refractive indices.
}
\label{fig:n_vs_E}
\end{figure*}

Note that the fitted values of the zero-field
refractive indices are: $n_h\approx 1.67$ and $n_p\approx 1.57$.
Figure~\ref{fig:n_vs_E} shows how the principal refractive indices,
$n_{\pm}$ and $n_z$,
change with applied electric field.
It is clear  that electric-field-induced optical anisotropy 
is weakly biaxial.

\section{Light modulation in DHFLC cells}
\label{sec:modulation}

In this section,
modulation of light in the DHFLC cells
studied in the previous section
will be of our primary concern.
For both the transmissive and reflective modes,
the DHFLC modulator is found to be
affected by the presence of 
amplitude modulation.
We study how this modulation influences
the transformation characteristics of 
a DHFLC spatial light modulator
operating as an axicon
producing a ring-shaped far field 
distribution of light.

\subsection{Amplitudes and Phases}
\label{subsec:phases-amplitudes}

Typically, in experiments
dealing with
the electro-optic response
of DHFLC cells,  
the transmittance of normally incident 
light passing through crossed polarizers
is measured as a function of the applied electric field.
In the case of normal incidence,
the transmission and reflection matrices can be easily obtained
from the results of Refs.~\cite{Kiselev:jpcm:2007,Kiselev:pra:2008,Kiselev:pre:2:2014}
in the limit of the wave vectors with 
vanishing tangential component, $k_p=0$. 
For our purposes, we shall need to write down the resultant expression 
for the transmission matrix
\begin{align}
&
  \label{eq:T-norm}
 \mvc{T}(\psi_{\dd})
=
\frac{t_{+}+t_{-}}{2}\,
\mvc{I}_2+\frac{t_{+}-t_{-}}{2}\,\
\mvc{Rt}(2 \psi_{\dd})
\bs{\sigma}_3,
\\
&
 \label{eq:t-pm}
  t_{\pm}=\frac{1-\rho_{\pm}^2}{%
1-\rho_{\pm}^2\exp(2in_{\pm}h)
}\exp(i n_{\pm} h),
\\
&
 \label{eq:rho-pm}
\rho_{\pm}=\frac{n_{\pm}/\mu-n_{\med}/\mu_{\med}}{%
n_{\pm}/\mu+n_{\med}/\mu_{\med}},
\end{align}
where
\begin{subequations}
  \label{eq:matrices}
\begin{align}
&
  \label{eq:unit-s3}
  \mvc{I}_2
=
\begin{pmatrix}
  1&0\\
0& 1
\end{pmatrix},
\quad
  \bs{\sigma}_3=
\begin{pmatrix}
  1&0\\
0& -1
\end{pmatrix},
\\
&
  \label{eq:rot-matrix}
\mvc{Rt}(2\psi_\dd)
\equiv
\begin{pmatrix}
  \cos2\psi_\dd&-\sin2\psi_\dd\\
\sin2\psi_\dd& \cos2\psi_\dd
\end{pmatrix}
\end{align}
\end{subequations}
and $h=k_{\vac} D$ is the thickness parameter.
Equation~\eqref{eq:T-norm} defines the transmission matrix 
linking the vector amplitudes of incident and transmitted waves,
$\vc{E}_0$ and $\vc{E}_{\ind{transm}}$,
through the standard input-output relation
\begin{align}
  \label{eq:input-output}
  \vc{E}_{\ind{transm}}=\mvc{T}(\psi_{\dd})\vc{E}_0,
\quad
  \vc{E}_{\ind{transm}}=\begin{pmatrix}
  E_x^{(\ind{transm})}\\
E_y^{(\ind{transm})}
\end{pmatrix}.
\end{align}

\begin{figure*}[!tbh]
\centering
  \resizebox{120mm}{!}{\includegraphics*{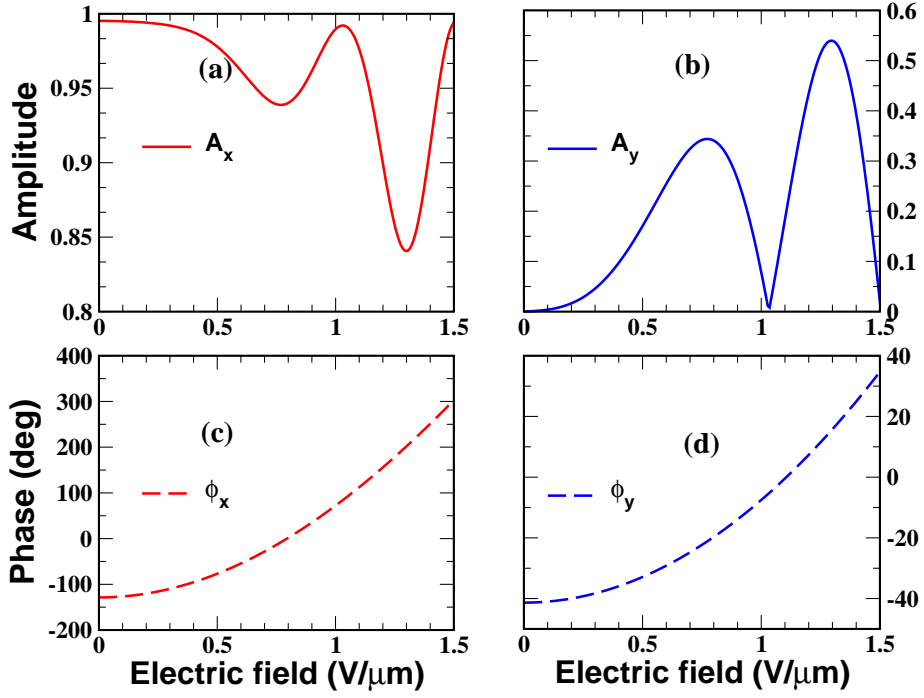}}
\caption{%
(Color online)
Amplitudes, $A_x$ and $A_y$ and phases, $\phi_x$ and $\phi_y$, 
for the components of the transmitted light vector amplitude
$\vc{E}_{\ind{transm}}/E_0$ computed
as a function of applied electric field in the transmissive mode. 
}
\label{fig:transm}
\end{figure*}

When the incident wave is linearly polarized along the
$x$ axis (the helix axis) with 
\begin{align}
  \label{eq:incident}
  \vc{E}_0=\begin{pmatrix}
  E_0\\
0
\end{pmatrix},
\end{align}
the components of the transmitted wave can be written
in the following form
\begin{align}
  \label{eq:transm}
&
  E_x^{(\ind{transm})}/E_0=A_x\exp(i\phi_x)=t_{+}-(t_{+}-t_{-})\sin^2\psi_\dd,
\notag
\\
&
E_y^{(\ind{transm})}/E_0=A_y\exp(i\phi_y)=\frac{t_{+}-t_{-}}{2}\sin (2\psi_\dd),
\end{align}
where $A_{x,\,y}$ and $\phi_{x,\, y}$ are the normalized amplitude and phase
of the corresponding complex amplitude component, respectively.
Then the transmittance coefficient
describing the intensity of the light passing through 
crossed polarizers is given by 
\begin{align}
&
  \label{eq:T-xy}
  |A_{y}|^2=\frac{|t_{+}-t_{-}|^2}{4}\,\sin^2(2\psi_\dd).
\end{align}

Note that, under certain conditions such as $|\rho_{\pm}|\ll 1$,
both the transmission coefficients~\eqref{eq:t-pm}
and the transmittance~\eqref{eq:T-xy} can be approximated by simpler formulas 
\begin{subequations}
  \label{eq:small-rho_approx}
\begin{align}
  \label{eq:t_pm_approx}
&
  t_{\pm}\approx \exp(i\Phi_{\pm}), 
\quad
\Phi_{\pm}=n_{\pm} h,
\\
\label{eq:tp-tm-approx}
&
\frac{t_{+}-t_{-}}{2}\approx
\sin(\Delta\Phi/2)\exp[i(\Phi_{+}+\Phi_{-}+\pi)/2],
\\
&
\label{eq:T-xy-approx}
 |A_{y}|^2 \approx\sin^2(\Delta\Phi/2)\,\sin^2(2\psi_\dd), 
\end{align}
\end{subequations}
where
$\Delta\Phi=\Phi_{+}-\Phi_{-}$
is the difference in optical path of the ordinary and extraordinary
waves known as the \textit{phase retardation}.

The complex valued components of the transmitted
plane wave are thus characterized by the
amplitudes and phases, $A_{x,\,y}$ and $\phi_{x,\,y}$,
given in Eq.~\eqref{eq:transm}.
These can now be computed
as a function of the applied electric field
by using 
our experimental data combined with
the results of fitting presented
in Sec.~\ref{subsec:results}.
Referring to Fig.~\ref{fig:transm},
rotation of the optical axis
combined with electric-field-induced  
change in the phase retardation
produces variations of
the amplitudes $A_x$ and $A_y$ with electric field.

For the component parallel to the helix axis, $E_x$,
the amplitude modulation,
however,
has negligibly small effect on 
the electric field dependence of the phase $\phi_x$.
It turned out that, despite amplitude modulation, 
this phase is close to $\Phi_{+}$:
$\phi_x\approx\Phi_{+}$.

From Eq.~\eqref{eq:T-xy-approx},
it might be concluded that
the phase of the component perpendicular to the helix axis,
$\phi_y$ is given by $(\Phi_{+}+\Phi_{-}+\pi)/2$.
It should be noted that,
in the approximation described by Eq.~\eqref{eq:t_pm_approx},
when
the factor $\sin(\Delta\Phi/2)$ changes its sign
at the points where the amplitude $A_y$ (and $T_{xy}$) 
is zero,
the phase should experience jump-like behaviour 
with $\Delta\phi_{x}=\pm\pi$.
These jumps are not shown in Fig.~\ref{fig:transm}
as, in real experiments, the amplitude $A_y$
never reaches zero due to scattering effects.

\begin{figure*}[!tbh]
\centering
  \resizebox{120mm}{!}{\includegraphics*{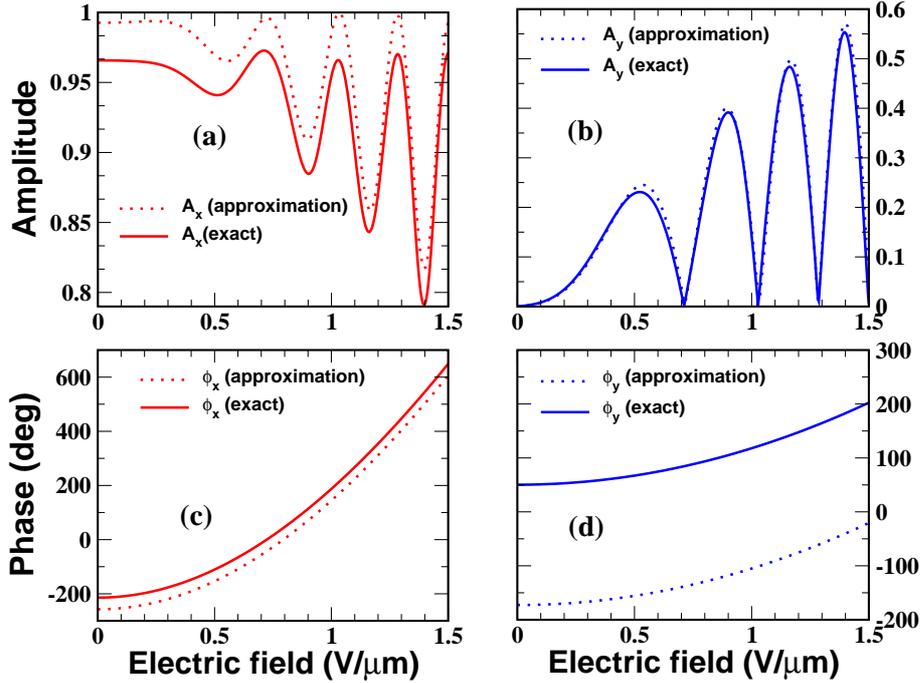}}
\caption{%
(Color online)
Amplitudes, $A_x$ and $A_y$ and phases, $\phi_x$ and $\phi_y$, 
for the components of the reflected light vector amplitude
$\vc{E}_{\ind{refl}}/E_0$ computed
as a function of applied electric field in the reflective mode.
Dotted lines represent the curves computed
from Eq.~\eqref{eq:transm} with
the doubled layer thickness
($D$ is replaced by $2D$).
Solid curves are evaluated using the reflection
matrix~\eqref{eq:R-norm}
and Eq.~\eqref{eq:refl}. 
}
\label{fig:refl}
\end{figure*}

We have also computed the amplitudes and phases
of the reflected wave
for the DHFLC cell operating in the reflective mode
with the mirror placed at the exit face of the cell.
It can be shown that, for the case of normal incidence, 
the reflection matrix can be written as follows
(details are relegated to Appendix~\ref{sec:reflection-matrix})
\begin{align}
&
  \label{eq:R-norm}
 \mvc{R}(\psi_{\dd})
=
\frac{r_{+}+r_{-}}{2}\,
\bs{\sigma}_3
+\frac{r_{+}-r_{-}}{2}\,\
\mvc{Rt}(-2 \psi_{\dd})
\\
&
 \label{eq:r-pm}
  r_{\pm}=\frac{\rho_{\pm}+\tilde{\rho}_{\pm}\exp(2i\Phi_{\pm})}{%
1+\rho_{\pm}\tilde{\rho}_{\pm}\exp(2i\Phi_{\pm})
},
\quad
\tilde{\rho}_{\pm}=\frac{R_r-\rho_{\pm}}{%
1-R_r\rho_{\pm}},
\end{align}
where
$R_r$ is the reflection coefficient of the mirror
given by
\begin{align}
  \label{eq:R_r}
  R_r=\rho_r \frac{1-\exp(2i n_r h_r)}{1-\rho_r^2\exp(2i n_r h_r)},
\quad
\rho_{r}=\frac{n_{r}/\mu_r-n_{\med}/\mu_{\med}}{%
n_{r}/\mu_r+n_{\med}/\mu_{\med}},
\end{align}
$n_r$ ($\mu_r$) and $h_r=k_\vac D_r$ ($D_r$) is 
the mirror refractive index (magnetic permittivity)
and the thickness parameter (thickness)
of the mirror, respectively.

In the reflective mode,
for the incident wave~\eqref{eq:incident},
the input-output relation
\begin{align}
  \label{eq:input-output}
  \vc{E}_{\ind{refl}}=\mvc{R}(\psi_{\dd})\vc{E}_0,
\quad
  \vc{E}_{\ind{refl}}=\begin{pmatrix}
  E_x^{(\ind{refl})}\\
E_y^{(\ind{refl})}
\end{pmatrix}
\end{align}
gives the components
of the reflected wave
\begin{align}
  \label{eq:refl}
&
  E_x^{(\ind{refl})}/E_0=A_x\exp(i\phi_x)=r_{+}-(r_{+}-r_{-})\sin^2\psi_\dd,
\notag
\\
&
E_y^{(\ind{refl})}/E_0=A_y\exp(i\phi_y)=-\frac{r_{+}-r_{-}}{2}\sin (2\psi_\dd).
\end{align}

The curves depicted in Fig.~\ref{fig:refl}
as solid lines are computed for the silver mirror
of the thickness $D_r=0.15$~\mum\ 
which is characterized by the complex valued refractive index~\cite{Rakic:aplopt:1998} 
$n_r\approx 0.16+3.8 i$ at the wavelength $\lambda=633$~nm.
The magnitude and the phase of the reflection
coefficient~\eqref{eq:R_r},
$R_r=|R_r|\exp(i\phi_r)$,
then can be estimated at about $|R_r|\approx 0.97$
and $\phi_r\approx 43$\dega.

Alternatively, the reflective mode
can  be described
in the double-layer approximation
using the transmission matrix~\eqref{eq:T-norm}
where the thickness of the cell is doubled
and $D$ is replaced by $2D$.
The results computed in this approximation
are shown in Fig.~\ref{fig:refl} as dotted lines.
Interestingly, for the amplitude modulation
of the component $A_y$,
the exact and approximated results
are in excellent agreement.
By contrast, the curves for the magnitude of the component 
$A_x$ [see Fig.~\ref{fig:refl}(a)] 
noticeably differ from each other.
It comes as no surprise that
the double-layer approximation, 
which is based on the assumption of perfect
reflection, overestimates $|A_x|$.

Similar to the transmissive mode,
the results for the phases, $\phi_x$ and
$\phi_y$, can be easily understood
in the limiting case where 
$|\rho_{\pm}|$ are small
and the reflection coefficients
$r_{\pm}$ defined in Eq.~\eqref{eq:r-pm} 
are close to $R_r\exp(2i\Phi_{\pm})$.
From Eq.~\eqref{eq:refl}, it can be inferred
that $\phi_x\approx 2\Phi_{+}+\phi_r$
and $\phi_y\approx \Phi_{+}+\Phi_{-}+\phi_r+3\pi/2$,
whereas the double-layer approximation
gives the relations: $\phi_x\approx 2\Phi_{+}$
and $\phi_y\approx \Phi_{+}+\Phi_{-}+\pi/2$.
The curves plotted in Figs.~\ref{fig:refl}(c)-(d)
clearly show
the phase shift introduced by the mirror
and the minus sign on the right hand side
of equation~\eqref{eq:refl} for 
the component $A_y$.
It can also be seen
that phase modulation
is about two times larger as compared to the case
of the transmissive mode.
So, the electric field required to reach
$2\pi$ modulation is reduced by the factor about $1.4$.

\subsection{Axicon}
\label{subsec:axicon}

The liquid crystal spatial light modulators are extensively used
for formation of light wavefields with a specified spatial 
distribution of intensity.
The above discussed effects of amplitude modulation 
may affect quality of the resulting light field.
In order to estimate how these effects influence
a reflective DHFLC modulator, 
we have modelled the DHFLC-SLM
as a 2D array of $256\times 256$ pixels
each of the area $200\times 200$~\mum$^2$.
It should be noted
that in our modelling
the electric field is assumed to be uniform.
In our case,
the pixel size to the cell thickness ratio 
is large enough 
for this assumption to be justified.

\begin{figure*}[!tbh]
\centering
\subfloat[Distribution of $\phi_{x}$]{
  \resizebox{70mm}{!}{\includegraphics*{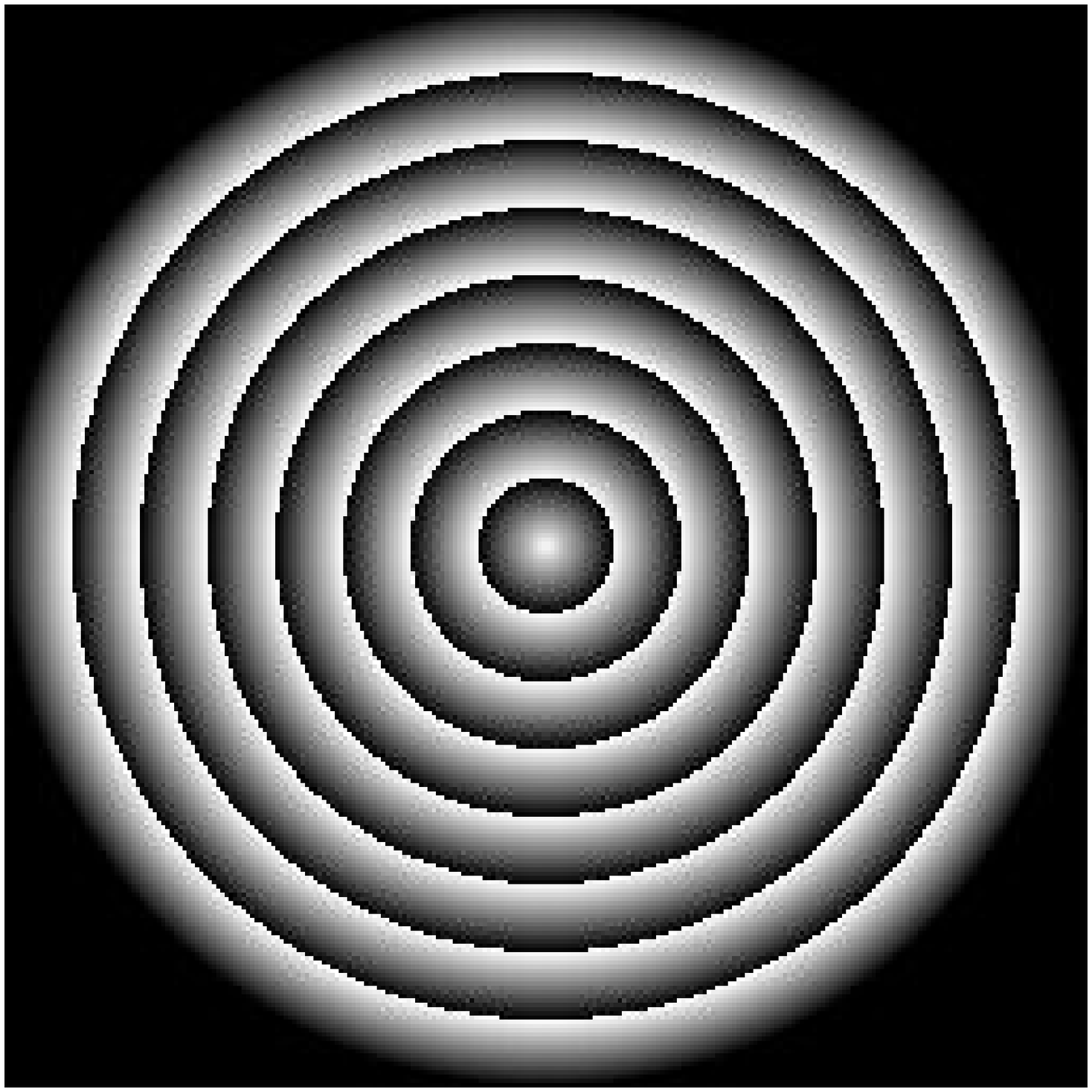}}
\label{subfig:phi_x}
}
\subfloat[Distribution of $A_{x}$]{
  \resizebox{70mm}{!}{\includegraphics*{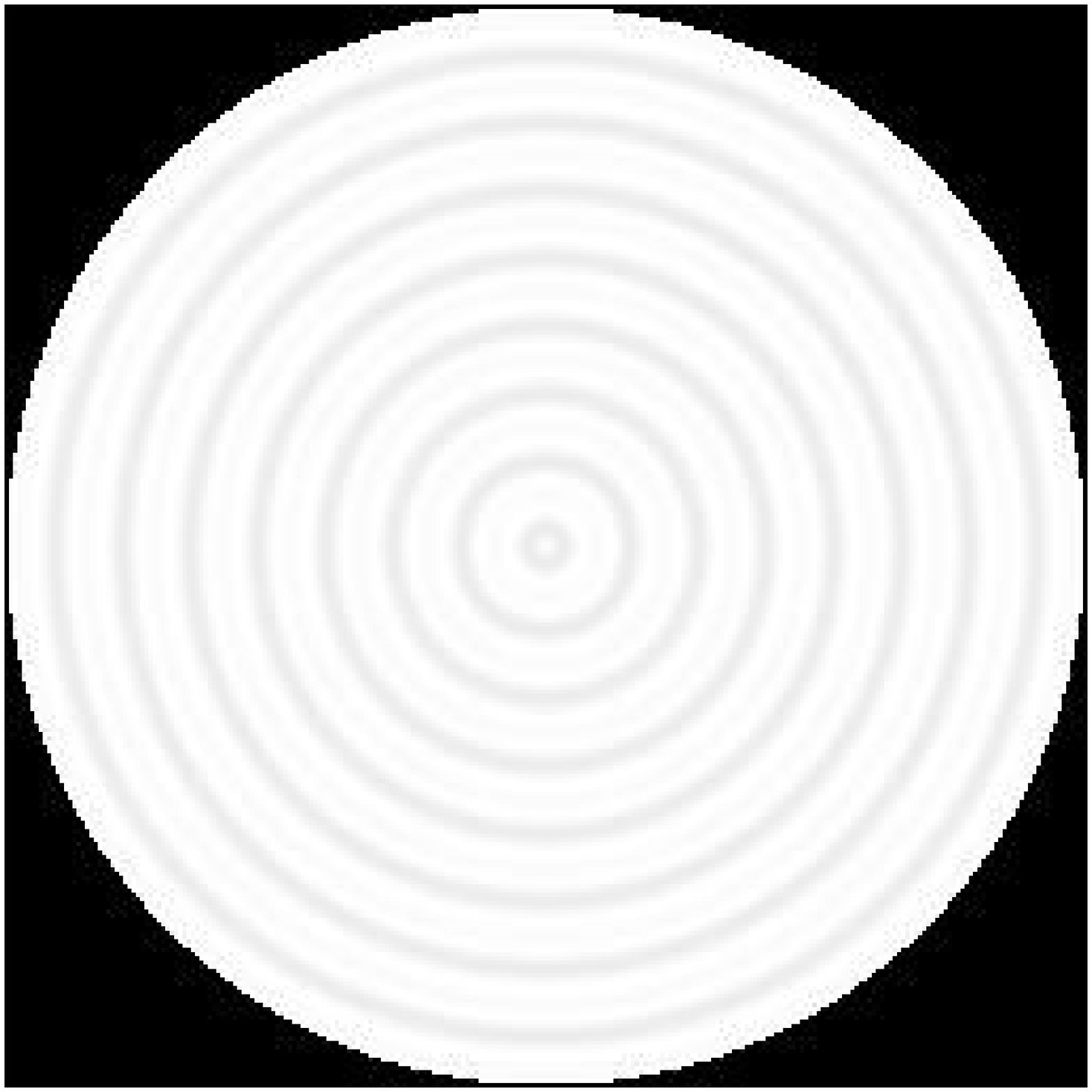}}
\label{subfig:A_x}
}
\caption{%
Profiles of (a)~phase, $\phi_x$, and (b)~amplitude, $A_x$, 
in the transverse plane of the DHFLC spatial light modulator.
Grayscale with $256$ levels of gray encodes the values of
$\phi_x$ and $A_x$ ranged from zero
(black color) to $2\pi$ and unity
(white color), respectively.
}
\label{fig:distribs}
\end{figure*}

In our calculations,
we assumed that the incident wave is linearly
polarized along the helix axis ($x$ axis)
and the phase distribution in the plane of the modulator
is equivalent to the phase profile of an axicon given by
\begin{align}
  \label{eq:phase_lens}
  \phi_x(x_0,y_0)=
-\Phi_{\ind{axicon}}\frac{\rho_0}{R},
\quad
\rho_0=\sqrt{x_0^2+y_0^2},
\end{align}
where $k=2\pi/\lambda$ is the wavenumber; 
$f$ is the focal length;
$R$ is the axicon radius
and $\Phi_{\ind{axicon}}$ is the phase modulation depth of the axicon. 
The $x$ component of the light field in the focal
plane of lens 
can be computed from 
the Fresnel-Kirchhoff diffraction
formula taken in the far field (Fraunhofer) approximation
(see, e.g., Ref.~\cite{Goodman:bk:2015})
\begin{align}
&
  \label{eq:E_x_focal}
  E_{x}(x,y)=\frac{\exp[ikf+\frac{i k(x^2+y^2)}{2f}]}{i\lambda f}
\int_{-\infty}^{\infty}\int_{-\infty}^{\infty}
A_x(x_0,y_0)
\notag
\\
&
\times
\exp\Bigl[
i\phi_x(x_0,y_0)
-\frac{i k}{f}
\left\{ x x_0+ y y_0\right\}
\Bigr]
\dd x_0\dd y_0.
\end{align}

The amplitude and phase distributions,
$A_x(x_0,y_0)$ and $\phi_x(x_0,y_0)$
that enter the integrand on the right hand side of
Eq.~\eqref{eq:E_x_focal} are approximated as follows:
(a)~we recast the phase profile of an axicon
characterized by the specified depth of modulation
into the step-like form with
the step height equal to $2\pi$;
(b)~the spatial phase distribution is then discretized 
along the coordinates $x$ and $y$ with
$\Delta x =\Delta y=200$~\mum\ 
so that each pixel is characterized by
the constant phase equal to the value 
of $\phi_x$ in its center;
(c)~for each value of the phase $\phi_x$,
we compute the corresponding value of
the amplitude $A_x$ and
derive the discretized distribution of the amplitude.
The final step involves 
using the standard FFT (fast Fourier transform)
technique~\cite{Brigham:bk:1988}
to evaluate of the integral on the right hand side 
of Eq.~\eqref{eq:E_x_focal}.

\begin{figure*}[!tbh]
\centering
\subfloat[Intensity in focal plane of the DHFLC axicon]{
  \resizebox{70mm}{!}{\includegraphics*{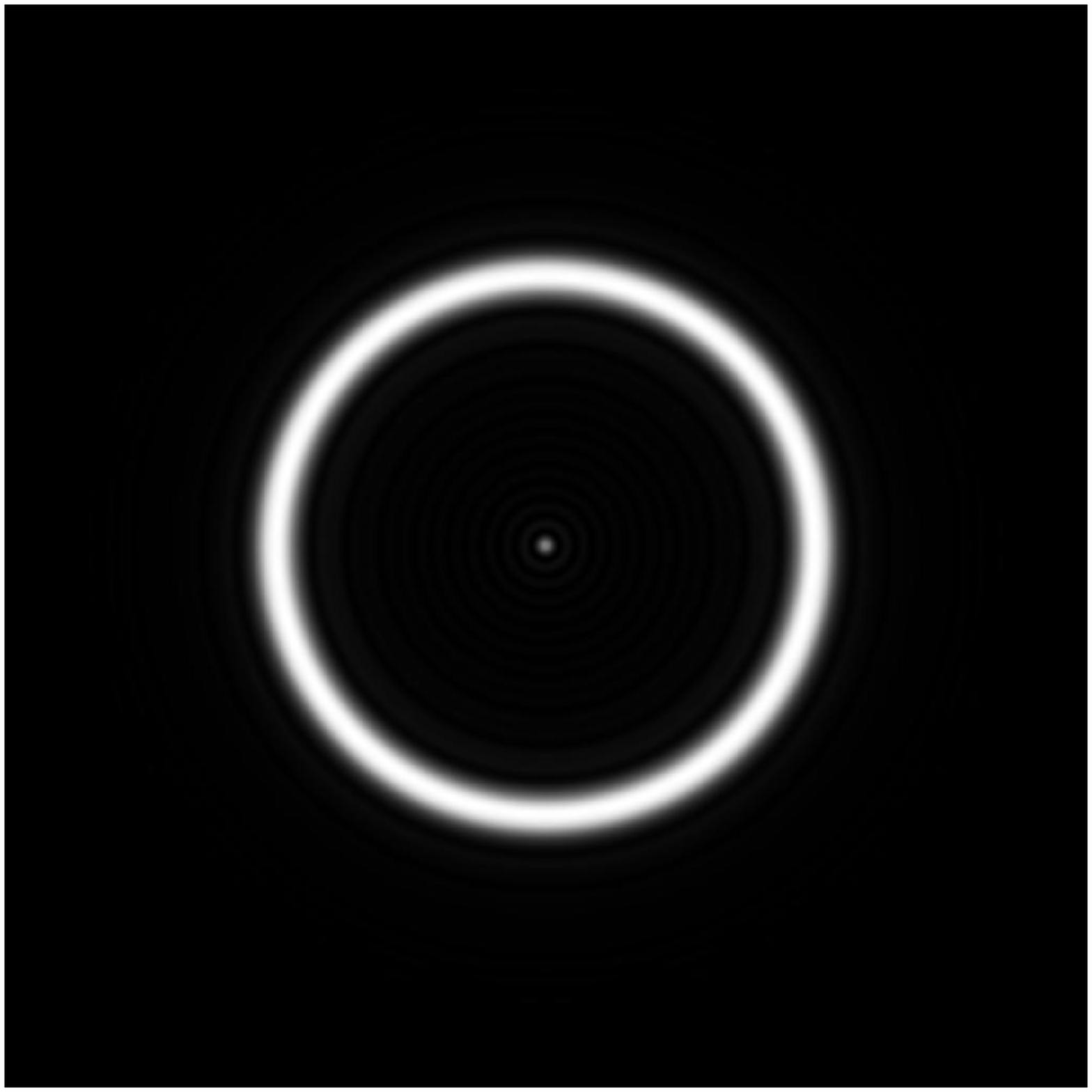}}
\label{subfig:focus-DHFLC}
}
 \subfloat[Intensity in focal plane of the ideal axicon]{
  \resizebox{70mm}{!}{\includegraphics*{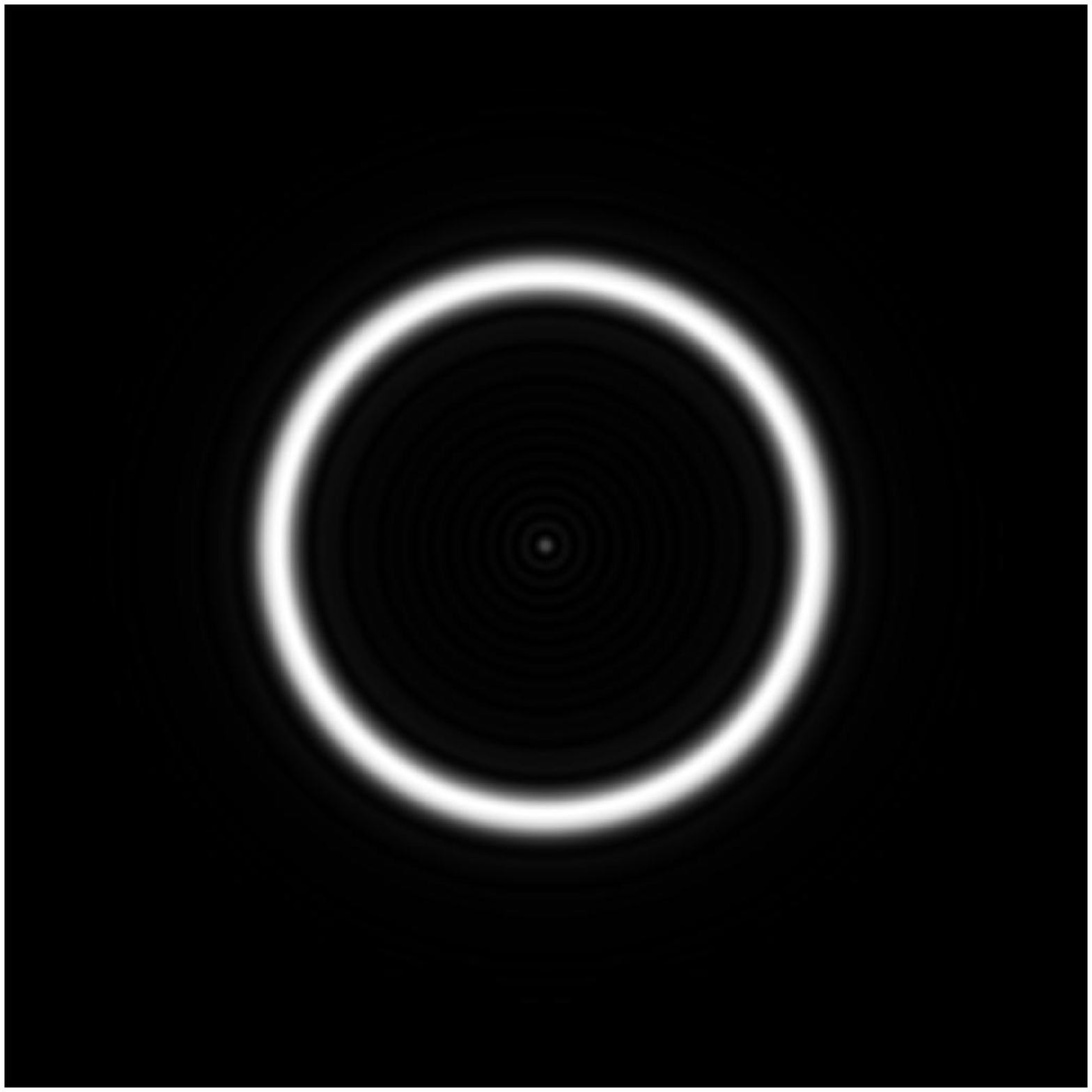}}
\label{subfig:focus-ideal}
}
\caption{%
Grayscale representation of the intensity distribution
in the focal plane
of (a)~the DHFLC
and (b)~ideal (no amplitude modulation) 
axicons operating in the reflective mode.
The cell thickness is about $50$~\mum.
}
\label{fig:intensity_focus}
\end{figure*}

\begin{figure*}[!tbh]
\centering
  \resizebox{120mm}{!}{\includegraphics*{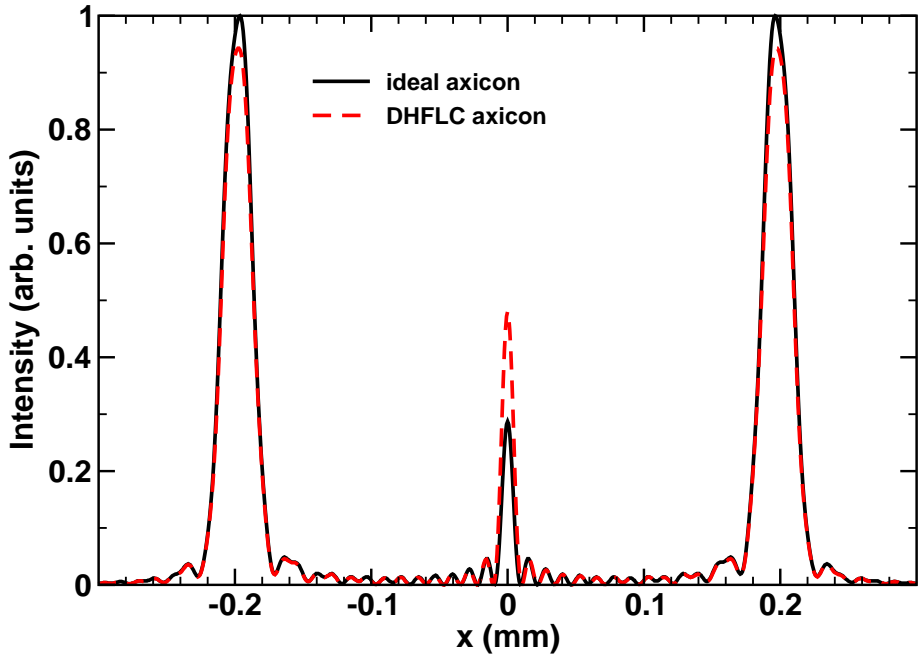}}
\caption{%
(Color online)
Intensity in the focal plane
as a function of $x$
for the DHFLC and ideal operating in the reflective mode.
Solid line represents the distribution computed
for the case of an ideal axicon without amplitude modulation.
The cell thickness is about $50$~\mum.
}
\label{fig:intensity_focus_x}
\end{figure*}

Figure~\ref{fig:distribs}
shows the phase and amplitude distributions
that were computed
for the modulation depth, $\Phi_{\ind{axicon}}$, 
equal to $16\pi$. 
The intensity distribution in the focal
plane 
and the corresponding $x$-dependence of the intensity
are presented in 
Fig.~\ref{fig:intensity_focus} and in Fig.~\ref{fig:intensity_focus_x},
respectively.
In Fig.~\ref{fig:intensity_focus_x},
the case of DHFLC axicon (dashed line)
(see Fig.~\ref{subfig:focus-DHFLC})
is compared with the curve (solid line)
computed for an ideal axicon without amplitude
modulation (see Fig.~\ref{subfig:focus-ideal}). 
The latter implies that
the amplitude $A_x(x_0,y_0)$ is assumed to be constant.
For the curves depicted in Fig.~\ref{fig:intensity_focus_x},
loss of power caused by amplitude modulation 
can be estimated to be below 5\%.

Note that the results shown in Figs~\ref{fig:distribs}~--~\ref{fig:intensity_focus_x},
are computed at the DHFLC cell thickness
taken to be about $50$~\mum.
We have also found that, when
the thickness of the DHFLC cells is halved, 
reducing the thickness
results in an increase of the power loss
that can be estimated at about 11\%.

\section{Discussion and conclusions}
\label{sec:discussion}

In this paper,
we have studied 
optical properties of the planar aligned
DHFLC cells 
(geometry is shown in Fig.~\ref{fig:geom})
that govern
modulation of light in such cells.
In our experiments, 
we have combined 
the experimental data for
electric field dependence
of in-plane optical axis orientation 
characterized by the azimuthal angle
$\psi_\dd$
(see Fig.~\ref{fig:phi_expt_th})
with the experimental technique
based on the Mach-Zehndler interferometer
to measure
the principal refractive indices
as a function of the electric field
(see Fig.~\ref{fig:delta-n}).

General expression for the effective optical tensor
of a DHFLC cell 
derived in Ref.~\cite{Kiselev:pre:2:2014}
[see Eq.~\eqref{eq:eff-diel-planar}]
was used to fit
the experimental curves.
The obtained parameters of 
the FLC mixture FLC-587
show  that this material is a weakly biaxial FLC.
Similarly, 
the resulting electric field dependence of
the principal effective refractive indices
of the DHFLC cell plotted in Fig.~\ref{fig:n_vs_E}
clearly indicate 
electric-field-induced optical biaxiality
of the effective dielectric tensor~\eqref{eq:eff-diel-planar}.
Note that optical biaxiality of the smectic C$^*$ phase
was previously reported in 
Refs~\cite{Gorec:jjap:1990,Song:pre:2:2007,Song:josa:2008}.

We have analyzed
modulation of light in the DHFLC cell under consideration
by computing the amplitudes and phases of the components
of light wave transmitted through (reflected from)
the cell in the transmissive (reflective) mode
described using the transfer matrix approach~\cite{Kiselev:pre:2:2014}.
It is found that, in addition to phase modulation,
electric-field-induced birefringence and rotation of
the in-plane optical axes generally result in
the effects of amplitude modulation.
Additionally,
we have calculated
the far field intensity distribution formed by
the DHFLC spatial light modulator
operating as an axicon.
From a comparison between the results
obtained for the ideal (no amplitude modulation) axicon
and the DHFLC modulator,
it can be concluded that
the characteristics of the DHFLC axicon
are very close to the ones of the ideal axicon.
In particular, 
for the cell thickness about $50$~\mum\  
the power loss due to amplitude modulation
is estimated at about 5\%,
whereas it increases up to 11\% at
the halved thickness.
For many applications,
such values of the power loss 
can be regarded as acceptable.
Note that, according to Ref.~\cite{Kotova:bull:2008}, 
the vortex lightwave fields
shaped in the form of certain curves,
which are of importance for efficient laser micromanipulation,
can be generated using 
the 2D array made up of, at least, 
$128\times 128$ amplitude-phase diffractive
elements.

In addition, the typical response time of the DHFLC modulator
based on FLC-587 is known to be around $100$~$\mu$s 
at the phase shift equal 
to $\pi$~\cite{Pozhidaev:jsid:2012}.
So, it might be concluded that
all essential prerequisites are in place 
for elaboration of the DHFLC spatial
light modulator operating almost as the ideal axicon 
at the phase modulation frequency about 1~kHz.

\begin{acknowledgments}
This work is supported by the 
the RFBR project No.~13-02-00598\_a.
\end{acknowledgments}

\appendix

\section{Effective dielectric tensor}
\label{sec:diel-tensor-dhf}

According to Ref.~\cite{Kiselev:pre:2011},
the effective dielectric tensor
\begin{align}
  \label{eq:eff-diel-tensor}
&
\bs{\varepsilon}_{\eff}=
\begin{pmatrix}
  \epsilon_{xx}^{(\eff)} & \epsilon_{xy}^{(\eff)}& \epsilon_{xz}^{(\eff)}\\
\epsilon_{yx}^{(\eff)}& \epsilon_{yy}^{(\eff)} & \epsilon_{yz}^{(\eff)}\\
\epsilon_{zx}^{(\eff)} & \epsilon_{zy}^{(\eff)} & \epsilon_{zz}^{(\eff)}
\end{pmatrix}  
\end{align}
can be expressed in terms of the averages
    \begin{align}
&
   \eta_{zz}=
\avr{\epsilon_{zz}^{-1}}
=
\epsilon_{0}^{-1}
\avr{[1+u_1 d_z^2+u_2 p_z^2]^{-1}},
\label{eq:eta}
\\
&
    \beta_{z\alpha}=
\avr{\epsilon_{z\alpha}/\epsilon_{zz}}
=
\left\langle
\frac{u_1 d_z d_\alpha+u_2 p_z p_\alpha}{1+u_1 d_z^2+u_2 p_z^2}
\right\rangle,
\label{eq:beta}
  \end{align}
where
$\avr{\ldots}\equiv\avr{\ldots}_{\phi}=(2\pi)^{-1}\int_{0}^{2\pi}\ldots\dd\phi$
and $\alpha\in\{x,y\}$,
as follows
\begin{align}
&
\epsilon_{zz}^{(\eff)}
=1/\eta_{zz},
\quad
\epsilon_{z\alpha}^{(\eff)}
=\beta_{z\alpha}/\eta_{zz},
\notag
\\
&
\epsilon_{\alpha\beta}^{(\eff)}
=\avr{\epsilon_{\alpha\beta}^{(P)}}+
\beta_{z\alpha} \beta_{z\beta}/\eta_{zz},
\label{eq:elements-eff-diel-tensor}
  \end{align}
where
$\avr{\epsilon_{\alpha\beta}^{(P)}}$
are the components of 
the averaged tensor,
$\avr{\bs{\varepsilon}_P}$:
\begin{align}
&
\avr{\epsilon_{\alpha\beta}^{(P)}}=
\left\langle
 \epsilon_{\alpha\,\beta}-\frac{ \epsilon_{\alpha\,z}
   \epsilon_{z\,\beta}}{ \epsilon_{zz}}
\right\rangle
\notag
\\
&
=
\epsilon_{0}
\left\langle
\delta_{\alpha\beta}+\frac{u_1 d_\alpha d_\beta+u_2 p_\alpha
  p_\beta+u_1 u_2 q_\alpha q_\beta}{1+u_1  d_z^2+u_2  p_z^2}\,
\right\rangle,
\label{eq:in-plane}
\\
&
q_{\alpha}=p_z d_{\alpha}-d_z p_{\alpha},
\quad 
\alpha,\beta\in\{x,y\},
\label{eq:q-vector}
\end{align}
describing effective in-plane anisotropy
that governs propagation of normally incident plane waves.

General
formulas~\eqref{eq:eta}-\eqref{eq:q-vector}
give  the zero-order approximation
for homogeneous models
describing the optical properties of
short-pitch DHFLCs~\cite{Kiselev:pre:2011,Kiselev:pre:2013}.

\section{Derivation of the reflection matrix}
\label{sec:reflection-matrix}

In this Appendix,
our task is to derive the reflection matrix~\eqref{eq:R-norm}
for the plane wave normally incident on
the DHFLC cell which exit face is covered with
a thin reflecting layer (mirror).
For this purpose, we shall use the transfer matrix
approach in the form presented in
Refs.~\cite{Kisel:pra:2008,Kiselev:jetp:2010,Kiselev:pre:2:2014}.
We begin with the $4\times 4$ transfer matrix
\begin{align}
  \label{eq:transfer-matr-full}
  \mvc{W}=
  \begin{pmatrix}
    \mvc{W}_{11} & \mvc{W}_{12}\\
    \mvc{W}_{21} & \mvc{W}_{22}
  \end{pmatrix}
=\mvc{W}_{\ind{FLC}}\mvc{W}_r,
\end{align}
where $\mvc{W}_{ij}$ are $2\times 2$ block-matrices,
expressed as a product of the transfer matrices
of the FLC cell and the mirror, $\mvc{W}_{\ind{FLC}}$
and $\mvc{W}_r$, respectively.
Given the transfer matrix~\eqref{eq:transfer-matr-full},
the transmission and reflection matrices,
$\mvc{T}$ and $\mvc{R}$,
can generally be computed from the formulas
(see, e.g., Ref.~\cite{Kiselev:pre:2:2014})
\begin{align}
  \label{eq:TR_W}
  \mvc{T}=\mvc{W}_{11}^{-1},
\quad
\mvc{R}=\mvc{W}_{21}\mvc{W}_{11}^{-1}
\end{align}
that relate $\mvc{T}$ and $\mvc{R}$
with the block matrices $\mvc{W}_{11}$
and $\mvc{W}_{21}$.

Applying Eq.~\eqref{eq:TR_W}
to the case of light normally impinging upon 
the FLC layer characterized by the
dielectric tensor~\eqref{eq:eff-diel-diag-planar}
gives the transmission matrix
\begin{align}
&
  \label{eq:T_FLC}
  \mvc{T}_{\ind{FLC}}=
[\mvc{W}_{11}^{(\ind{FLC})}]^{-1}
\notag
\\
&
=
\mvc{Rt}(\psi_{\ind{d}})
\begin{pmatrix}
  t_{+}^{(\ind{FLC})} & 0\\
0 & t_{-}^{(\ind{FLC})}
\end{pmatrix}
\mvc{Rt}(-\psi_{\ind{d}}),
\\
&
 \label{eq:t-pm-FLC}
  t_{\pm}^{(\ind{FLC})}=\frac{1-\rho_{\pm}^2}{%
1-\rho_{\pm}^2\exp(2in_{\pm}h)
}\exp(i n_{\pm} h),
\\
&
 \label{eq:rho-pm-FLC}
\rho_{\pm}=\frac{n_{\pm}/\mu-n_{\med}/\mu_{\med}}{%
n_{\pm}/\mu+n_{\med}/\mu_{\med}},
\end{align}
where $\mvc{Rt}(\psi_{\ind{d}})$
is the rotation matrix [see Eq.~\eqref{eq:rot-matrix}],
which is identical to the matrix given in Eq.~\eqref{eq:T-norm}.
The corresponding result for the reflection matrix reads
\begin{align}
&
  \label{eq:R_FLC}
  \mvc{R}_{\ind{FLC}}=
\mvc{W}_{21}^{(\ind{FLC})}[\mvc{W}_{11}^{(\ind{FLC})}]^{-1}
\notag
\\
&
=
  \bs{\sigma}_3\mvc{Rt}(\psi_{\ind{d}})
\begin{pmatrix}
  r_{+}^{(\ind{FLC})} & 0\\
0 & r_{-}^{(\ind{FLC})}
\end{pmatrix}
\mvc{Rt}(-\psi_{\ind{d}}),
\\
&
 \label{eq:r-pm-FLC}
  r_{\pm}^{(\ind{FLC})}=\rho_{\pm}\frac{1-\exp(2in_{\pm}h)}{%
1-\rho_{\pm}^2\exp(2in_{\pm}h)
}.
\end{align}

For the reflecting layer, the results 
\begin{subequations}
  \label{eq:TR_r}
\begin{align}
&
  \label{eq:T_r}
  \mvc{T}_{r}=
[\mvc{W}_{11}^{(r)}]^{-1}=
T_r\mvc{I}_2,
\\
&
  \label{eq:t_r}
  T_{r}=\frac{1-\rho_{r}^2}{%
1-\rho_{r}^2\exp(2in_{r}h_r)
}\exp(i n_{r} h_r),
\\
&
  \label{eq:R_r1}
\mvc{R}_{r}=
\mvc{W}_{21}^{(r)}[\mvc{W}_{11}^{(r)}]^{-1}=
  R_r \bs{\sigma}_3,
\\
&
  \label{eq:r_r}
  R_{r}=\rho_{r}\frac{1-\exp(2in_{r}h_r)}{%
1-\rho_{r}^2\exp(2in_{r}h_r)
},
\:
\rho_{r}=\frac{n_{r}/\mu_r-n_{\med}/\mu_{\med}}{%
n_{r}/\mu_r+n_{\med}/\mu_{\med}}.
\end{align}
\end{subequations}
can be obtained
from the above formulas for the FLC layer by replacing
$\{n_{+},n_{-},h=k_{\vac}D\}$ with
$\{n_{r},n_{r},h_r=k_{\vac}D_r\}$.

For a non-absorbing FLC material,
the unitarity relations
imply the symmetry conditions~\cite{Kiselev:pre:2:2014}.
\begin{align}
&
  \label{eq:unitarity-T}
  [\mvc{W}_{ii}^{(\ind{FLC})}]^{T}=\mvc{W}_{ii}^{(\ind{FLC})},
\quad
  [\mvc{W}_{21}^{(\ind{FLC})}]^{T}=-\mvc{W}_{12}^{(\ind{FLC})},
\\
&
\label{eq:unitarity-H-11}
  [\mvc{W}_{11}^{(\ind{FLC})}]^{\dagger}=\bs{\sigma}_3\mvc{W}_{22}^{(\ind{FLC})}\bs{\sigma}_3,
\\
&
\label{eq:unitarity-H-21}
  [\mvc{W}_{21}^{(\ind{FLC})}]^{\dagger}=-\bs{\sigma}_3\mvc{W}_{21}^{(\ind{FLC})}\bs{\sigma}_3,
\end{align}
where a dagger and the superscript $T$ will denote
Hermitian conjugation and
matrix transposition, respectively.
The conditions~\eqref{eq:unitarity-T} can now be used to express
the block-matrix $\mvc{W}_{11}$
in terms of the transmission and reflection matrices
as follows
\begin{align}
&
  \label{eq:W_11}
  \mvc{W}_{11}=\mvc{W}_{11}^{(\ind{FLC})}\mvc{W}_{11}^{(r)}+\mvc{W}_{12}^{(\ind{FLC})}\mvc{W}_{21}^{(r)}
\notag
\\
&
=[\mvc{W}_{11}^{(\ind{FLC})}+\mvc{W}_{12}^{(\ind{FLC})}\mvc{R}_{r}]\mvc{T}_r^{-1}
\notag
\\
&
=\mvc{T}_{\ind{FLC}}^{-1}
[\mvc{I}_2-\mvc{R}_{\ind{FLC}}^{T}\mvc{R}_{r}]\mvc{T}_r^{-1}.
\end{align}
Equation~\eqref{eq:W_11} combined with the relation~\eqref{eq:TR_W}
immediately give the transmission matrix
in the following general form
\begin{align}
  \label{eq:T-general}
  \mvc{T}=\mvc{T}_r
[\mvc{I}_2-\mvc{R}_{\ind{FLC}}^{T}\mvc{R}_{r}]^{-1}\mvc{T}_{\ind{FLC}}.
\end{align}
Similarly, using the symmetry relations~\eqref{eq:unitarity-H-11} and~\eqref{eq:unitarity-H-21},
we can obtain the following result for the block-matrix $\mvc{W}_{21}$:
\begin{align}
&
  \label{eq:W_21}
  \mvc{W}_{21}=\mvc{W}_{21}^{(\ind{FLC})}\mvc{W}_{11}^{(r)}+\mvc{W}_{22}^{(\ind{FLC})}\mvc{W}_{21}^{(r)}
\notag
\\
&
=[\mvc{W}_{21}^{(\ind{FLC})}+\mvc{W}_{22}^{(\ind{FLC})}\mvc{R}_{r}]\mvc{T}_r^{-1}
\notag
\\
&
=\bs{\sigma}_3[\mvc{T}_{\ind{FLC}}^{\dagger}]^{-1}
\bigl\{\bs{\sigma}_3\mvc{R}_{r}-[\bs{\sigma}_3\mvc{R}_{\ind{FLC}}]^{\dagger}\bigr\} \mvc{T}_r^{-1}.
\end{align}
Substituting formulas~\eqref{eq:W_11} and~\eqref{eq:W_21}
into Eq.~\eqref{eq:TR_W} gives  
the resulting expression for the reflection matrix
in the following form:
\begin{align}
&
  \label{eq:R-general}
  \mvc{R}=
\bs{\sigma}_3 [\mvc{T}_{\ind{FLC}}^{\dagger}]^{-1}
\bigl\{\bs{\sigma}_3\mvc{R}_{r}-[\bs{\sigma}_3\mvc{R}_{\ind{FLC}}]^{\dagger}\bigr\}
\notag
\\
&
\times\bigl[\mvc{I}_2-\mvc{R}_{\ind{FLC}}^{T}\mvc{R}_{r}\bigr]^{-1}\mvc{T}_{\ind{FLC}}.
\end{align}

Formulas~\eqref{eq:T-general} and~\eqref{eq:R-general} are quite
general. In particular, they can be applied to any
non-absorbing uniformly anisotropic layer represented by the FLC
material.
For the case of normal incidence,
from Eqs.~\eqref{eq:T_FLC}--~\eqref{eq:TR_r},
it follows that, similar to the matrices 
$\mvc{T}_{\ind{FLC}}(\psi_\dd)$ and $\mvc{R}_{\ind{FLC}}(\psi_\dd)$,
the matrices $\mvc{T}(\psi_\dd)$ and $\mvc{R}(\psi_\dd)$
can be recast into the following
factorized form:
\begin{subequations}
  \label{eq:TR-factorized}
\begin{align}
&
  \label{eq:T-factorized}
  \mvc{T}(\psi_\dd)=
\mvc{Rt}(\psi_\dd)\mvc{T}(0)\mvc{Rt}(-\psi_\dd),
\\
&
  \label{eq:R-factorized}
  \mvc{R}(\psi_\dd)=
\mvc{Rt}(-\psi_\dd)\mvc{R}(0)\mvc{Rt}(-\psi_\dd).
\end{align}
\end{subequations}
The matrices, $\mvc{T}(0)$ and $\mvc{R}(0)$,
are diagonal and can be easily computed.
After algebraic manipulations,
it is not difficult to derive the analytical expression
for the transmission matrix
given in Eqs.~\eqref{eq:R-norm}--\eqref{eq:R_r}.


%

\end{document}